\documentclass[twocolumn,showpacs,amsmath,amssymb,pra,superscriptaddress]{revtex4-1}
\usepackage{epsfig}
\usepackage{epstopdf}
\usepackage{graphicx}
\usepackage{dcolumn}
\usepackage{color}
\usepackage{bm}
\usepackage{mathrsfs}
\usepackage{bbold}
\usepackage{amssymb}
\usepackage{upgreek}
\DeclareMathOperator{\Tr}{Tr}

\begin{document}

\title{\Large Probing nonclassicality under spontaneous decay}

\author{Md.~Manirul Ali}
\email{mani@cts.nthu.edu.tw}
\affiliation{Physics Division, National Center for Theoretical Sciences, National Tsing Hua University,
Hsinchu 30013, Taiwan}
\author{Po-Wen Chen}
\email{powen@iner.gov.tw}
\affiliation{Physics Division, Institute of Nuclear Energy Research, Longtan, Taoyuan 32546, Taiwan}

\date{\today}

\begin{abstract}
We investigate the {\it nonclassicality} of an open quantum system using Leggett-Garg inequality (LGI)
which test the correlations of a single system measured at different times. Violation of LGI implies
nonclassical behavior of the open system. We investigate the violation of the Leggett-Garg inequality
for a two level system (qubit) spontaneously decaying under a general non-Markovian dissipative
environment. Our results are exact as we have calculated the two-time correlation functions exactly for
a wide range of system-environment parameters beyond Born-Markov regime.
\end{abstract}

\pacs{03.65.Yz, 03.67.Pp, 03.65.Ta}
\maketitle

\section{Introduction}\label{sec:introduction}

Quantum physics manifests nonclassical correlations through the violation of Bell and Leggett-Garg (LG) inequalities.
A renewed interest in the investigation of Leggett-Garg inequalities has gained momentum within past few years.
The original motivation of the seminal work \cite{Leggett85} by Leggett and Garg was to probe quantum coherence in
macroscopic systems \cite{Leggett02}. It is generally believed that the quantumness of a large system is destroyed by the many-body interactions
with a noisy environment, which is broadly termed as decoherence. Inspired by this fact, LG inequalities can play as an indicator
of nonclassicality for open quantum systems in a dissipative environment \cite{Breuer02}. Two classical assumptions are made in deriving the
LG inequalities ($a_1$){\it macrorealism}: macroscopic systems are always in a definite state with well-defined pre-existing
value, and ($a_2$) {\it noninvasive measurability}: this pre-existing value can be measured in a non-invasive way, that is,
without disturbing the subsequent dynamics of the system. Violation of  Leggett-Garg inequalities (LGI) in the quantum regime
indicate nonclassical behavior of the system due to the existence of superposition states violating assumption ($a_1$) and/or
due to the measurement induced collapse of state that violates assumption ($a_2$). The first experimental violation of LGI
was demonstrated in \cite{Palacios10} after which the LGI violation was probed in a diverse range of physical systems, for
example, photonic systems \cite{photons1,photons2,photons3,photons4}, nuclear magnetic resonance \cite{nmr1,nmr2},
phosphorus impurities in silicon \cite{silicon}, nitrogen-vacancy defect in diamond \cite{diamond}, and most recently in
superconducting flux qubit \cite{Knee}. Also, quantum violation of LGI has been studied theoretically in a variety of systems
like electrons in quantum dot \cite{Lambert1,Nathan,Ruskov}, optomechanical system \cite{Lambert2}, using quantum
nondemolition measurement applied to atomic ensemble \cite{QND}, oscillating neutral kaons and neutrino oscillations
\cite{dhome,PRL2016}, and even in biological light-harvesting protein complex \cite{biology1,biology2}.

Violation of LGI \cite{Lambert3} is associated to the nonclassical dynamics and measurement correlations in the quantum system.
Realistic quantum systems are open to unavoidable interaction with its surrounding environment that act as a source of decoherence and
dissipation, resulting to the loss of quantumness of the system. This loss of quantumness of the open quantum system can be
probed through the LG inequality that acts as witness of nonclassicality. In the past, the LGI violation has been discussed
for closed systems \cite{Leggett85,Leggett02,photons4,nmr1,nmr2,silicon,diamond,Knee,photons2,photons3,QND,dhome,PRL2016}.
The violations of the LG inequality was also investigated for open systems in the Born-Markov limit \cite{Palacios10,Lambert1,Nathan,Ruskov,Lambert2,photons1,biology1,biology2,Lambert3,YNC,Emary,Luczka}.
Earlier, we have investigated the violation of the Leggett-Garg inequality for a two level system under decoherence in a
non-Markovian dephasing environment \cite{NMAliChen}. Here we extend these results to a dissipative system-environment
coupling outside the Born-Markov regime. In the present paper, we consider a two-level system (qubit) spontaneously decaying
under a non-Markovian bosonic environment. The non-Markovian characteristic of the model is discussed elsewhere
\cite{nvtyJC1,nvtyJC2} in great detail. We briefly discuss the model and the method to calculate the non-Markovian
two-time correlation functions for the two-level system and its dynamical loss of quantumness through Leggett-Garg inequality.
Our analysis is exact and valid both for weak and strong system-environment couplings, as we have not performed neither the
Born nor the Markov approximation \cite{Scully97,Carmichael99}. Then we present our numerical results to investigate the
Leggett-Garg inequality in various system-environment parameter regime. Finally, a conclusion is given at the end.

\section{The model and exact two-time correlation functions}\label{sec:model}

We consider a two-level system (qubit) spontaneously decaying under an environment having a continuum
of modes. The total Hamiltonian of the system plus environment is given by
\begin{eqnarray}
\label{TotH}
H = H_{S} + H_{E} + H_{I}
\end{eqnarray}
where $H_S=\hbar \omega_0 \sigma_{+}\sigma_{-}$ describes the two-level system.
The operators $\sigma_{+} = |1\rangle_S\langle 0|$ and $\sigma_{-} = |0 \rangle_S\langle 1|$ with
ground state $|0\rangle_S$, excited state $|1\rangle_S$, and transition frequency $\omega_0$.
The environment Hamiltonian $H_E = \sum_k \hbar \omega_k b_k^{\dagger} b_k$, which
describes a collection of harmonic oscillators with Bosonic operators $b_k^{\dagger}$ and $b_k$.
The interaction Hamiltonian is given by
\begin{eqnarray}
\label{IntH}
H_{I} = \sum_k \left( g_k \sigma_{+} \otimes b_k + g_k^{\ast} \sigma_{-} \otimes b_k^{\dagger} \right).
\end{eqnarray}
Next, we go the interaction picture with respect to $H_0 = H_S + H_E$, the time evolution of the total
system-plus-environment state in the interaction picture
\begin{eqnarray}
\label{Schrod}
\frac{d}{dt} |{\tilde \Psi} (t)\rangle = -\frac{i}{\hbar} {\tilde H}_I(t)  |{\tilde \Psi} (t)\rangle,
\end{eqnarray}
where
\begin{eqnarray}
\label{IntH2}
{\tilde H}_I(t) = \hbar \left[ \sigma_{+} \otimes B(t) +  \sigma_{-} \otimes B^{\dagger}(t) \right]
\end{eqnarray}
is the interaction picture operator ${\tilde H}_I(t) = e^{\frac{i H_0 t}{\hbar}} H_I e^{-\frac{i H_0 t}{\hbar}}$
with $B(t) = \sum_k g_k b_k e^{i(\omega_0 - \omega_k) t}$. We start with an initial product
state $|\Psi(0)\rangle = \left( c_0 |0\rangle_{S} + c_1(0) |1\rangle_{S} \right) \otimes |0\rangle_{E}$, where
the environment is initially in the vacuum state $|0\rangle_E$. The interaction Hamiltonian conserves the total particle number,
the Schr\"{o}dinger equation generated by ${\tilde H}_I(t)$ will be confined to the subspace spanned by the vectors
$|0\rangle_{S} \otimes |0\rangle_{E}$, $|1\rangle_{S} \otimes |0\rangle_{E}$, and $|0\rangle_{S} \otimes |k\rangle_{E}$.
The exact time evolution of $|\Psi(0)\rangle$ is given by
\begin{eqnarray}
\label{tPsi}
\nonumber
|{\tilde \Psi}(t)\rangle &=&  c_0 |0\rangle_{S} \otimes |0\rangle_{E} + c_1(t) |1\rangle_{S} \otimes |0\rangle_{E} \\
&&{} + \sum_k c_k (t) |0\rangle_{S} \otimes |k\rangle_{E}
\end{eqnarray}
where $|k\rangle_{E}=b_k^{\dagger}|0\rangle_{E}$ is the state with one particle in mode $k$. Note that
the amplitude $c_0$ is constant in time because ${\tilde H}_I(t)  |0\rangle_{S} \otimes |0\rangle_{E}=0$.
Substituting $|{\tilde \Psi}(t)\rangle$ from Eq.~(\ref{tPsi}) into the Schr\"{o}dinger equation (\ref{Schrod}),
one can obtain an integrodifferential equation for $c_1(t)$ as
\begin{eqnarray}
\label{amp1N}
\frac{d}{dt} c_1(t) = - \int_{0}^{t} d\tau g(t-\tau) c_1(\tau),
\end{eqnarray}
where $g(t-\tau)= \langle 0| B(t) B^{\dagger}(\tau) |0\rangle_{E}$ is the two-time correlation function of the
reservoir and is given by
\begin{eqnarray}
\nonumber
g(t-\tau) &=& \sum_k |g_k|^2 e^{i(\omega_0-\omega_k)(t-\tau)} \\
&=&  \int d\omega J(\omega) e^{i(\omega_0 - \omega)(t-\tau)}.
\end{eqnarray}
Here $J(\omega)$ is the spectral density of the environment. The reduced density operator of the system in the
interaction picture ${\tilde \rho}_{S} (t) = \Tr_E \{ |{\tilde \Psi}(t)\rangle \langle {\tilde \Psi}(t)| \}$ is
determined by the function $c_1(t)$. We can calculate the exact two-time correlation function
$\langle \sigma_{+}(t_1) \sigma_{-}(t_2) \rangle_I$ in the interaction picture as follows
\begin{eqnarray}
\nonumber
&&{} \langle \sigma_{+}(t_1) \sigma_{-}(t_2) \rangle_I \\
\nonumber
&&{} =\Tr_{S \oplus E} \left( {\tilde U}^{\dagger}(t_1) \sigma_{+} {\tilde U}(t_1) {\tilde U}^{\dagger}(t_2) \sigma_{-} {\tilde U}(t_2) \rho_T(0) \right) \\
\nonumber
&&{} =\Tr_{S \oplus E} \left( \sigma_{+} {\tilde U}(t_1-t_2) \sigma_{-} {\tilde U}(t_2) |\Psi(0)\rangle \langle \Psi(0)|
{\tilde U}^{\dagger}(t_1) \right) \\
&&{} =\Tr_{S \oplus E} \left( \sigma_{+} {\tilde U}(t_1-t_2) \sigma_{-} |{\tilde \Psi}(t_2)\rangle \langle {\tilde \Psi}(t_1)| \right)
\label{Exact1}
\end{eqnarray}
where ${\tilde U}(t)$ is the unitary time evolution operator generated by the Hamiltonian ${\tilde H}_I(t)$ and
${\tilde U}(t_1\!-\!t_2)={\tilde U}(t_1) {\tilde U}^{\dagger}(t_2)$. In Eq.~(\ref{Exact1}), we substitute the time evolved
$|{\tilde \Psi}(t_2)\rangle$ using Eq.~(\ref{tPsi}) to finally obtain
\begin{eqnarray}
\langle \sigma_{+}(t_1) \sigma_{-}(t_2) \rangle_I = c_1(t_2) c_1^{\ast}(t_1),
\end{eqnarray}
where we have used $\sigma_{+}{\tilde U}(t_1\!\!-\!\!t_2) \sigma_{-} |{\tilde \Psi}(t_2)\rangle
=c_1(t_2) |1\rangle_{S} \otimes |0\rangle_{E}$. Another two-time correlation function
$\langle \sigma_{-}(t_1) \sigma_{+}(t_2) \rangle_I$ can also be obtained exactly as follows
\begin{eqnarray}
\nonumber
&&{} \langle \sigma_{-}(t_1) \sigma_{+}(t_2) \rangle_I \\
\nonumber
&&{} =\Tr_{S \oplus E} \left( {\tilde U}^{\dagger}(t_1) \sigma_{-} {\tilde U}(t_1) {\tilde U}^{\dagger}(t_2)
\sigma_{+} {\tilde U}(t_2) \rho_T(0) \right) \\
\nonumber
&&{} =\Tr_{S \oplus E} \left( \sigma_{-} {\tilde U}(t_1-t_2) \sigma_{+} {\tilde U}(t_2) |\Psi(0)\rangle \langle \Psi(0)|
{\tilde U}^{\dagger}(t_1) \right) \\
&&{} =\Tr_{S \oplus E} \left( \sigma_{-} {\tilde U}(t_1-t_2) \sigma_{+} |{\tilde \Psi}(t_2)\rangle \langle {\tilde \Psi}(t_1)| \right)
\label{Exact2}
\end{eqnarray}
Again by substituting $|{\tilde \Psi}(t_2)\rangle$ explicitly in Eq.~(\ref{Exact2}) and using the fact that
$\sigma_{-} {\tilde U}(t_1\!-\!t_2) \sigma_{+} |{\tilde \Psi}(t_2)\rangle = c_0 c_1(t_1-t_2) |0\rangle_{S} \otimes |0\rangle_{E}$,
one can show
\begin{eqnarray}
\langle \sigma_{-}(t_1) \sigma_{+}(t_2) \rangle_I = |c_0|^2 c_1(t_1-t_2)
\end{eqnarray}
Using the transformation ${\tilde U}(t)=U_0^{\dagger}(t)U(t)$ with $U_0(t)$ and $U(t)$ being the unitary time
evolution operators generated by $H_0$ and $H$ respectively, it is then straightforward to obtain the two-time
correlation functions in the usual Heisenberg picture
\begin{eqnarray}
\nonumber
&&{} \langle \sigma_{+}(t_1) \sigma_{-}(t_2) \rangle = \langle \sigma_{+}(t_1) \sigma_{-}(t_2) \rangle_I \exp \{-i\omega_0 (t_2 - t_1) \} \\
\label{ttc1}
&&{} = c_1(t_2) c_1^{\ast}(t_1) \exp \{-i\omega_0 (t_2 - t_1) \}
\end{eqnarray}
and
\begin{eqnarray}
\nonumber
&&{} \langle \sigma_{-}(t_1) \sigma_{+}(t_2) \rangle = \langle \sigma_{-}(t_1) \sigma_{+}(t_2) \rangle_I \exp \{i\omega_0 (t_2 - t_1) \} \\
\label{ttc2}
&&{} = |c_0|^2 c_1(t_1-t_2) \exp \{i\omega_0 (t_2 - t_1) \}
\end{eqnarray}

\section{Probing non-classicality using two-time correlation function}\label{sec:twotime}

Leggett-Garg inequalities test the correlations of a single system measured at different times for which
we need to calculate the two-time correlation functions ``$\langle O(t_j) O(t_i) \rangle$'' of an
observable $O$. We can construct the simplest LGI as follows. Consider the measurement of
an observable $O(t)$ of a two level system which is found to take a value $+1$ or $-1$,
depending on the system being in state $| + \rangle$ or $| - \rangle$. Now perform a series of
three set of experimental runs starting from identical initial condition (at time $t = 0$) such
that in the first set of runs $O$ is measured at times $t_1$ and $t_2 = t_1 + \tau$; in the second,
at $t_1$ and $t_3 = t_1 + 2 \tau$; in the third at $t_2$ and $t_3$ (where $t_3 > t_2 > t_1 $).
The temporal correlations $\langle O(t_j) O(t_i) \rangle$ can be obtained from such measurements.
Leggett and Garg \cite{Leggett85} followed the standard classical argument (assumptions $a_1$ and $a_2$)
leading to a Bell-type inequality, with times $t_i$ and $t_j$ playing the role of apparatus settings.
According to the classical assumption $a_1$, for any set of runs corresponding to the same initial state, any individual
$O(t)$ has a well-defined pre-existing value prior to measurement. According to assumption $a_2$, the value of
$O(t_j)$ or $O(t_i)$ in any pair does not depend on whether any prior or subsequent measurement has been made
on the system, so the joint measurements $O(t_j) O(t_i)$ are independent of the sequence in which they are measured.
Hence for classical systems the combination $O(t_2) O(t_1) + O(t_3) O(t_2) - O(t_3) O(t_1)$
has an upper bound of $+1$ and lower bound of $-3$. Replacing all the individual product terms
in this expression by their averages over the entire ensemble for each sets of runs, one obtains the
following form of LGI
\begin{eqnarray}
C_3 = C_{21} + C_{32} - C_{31} \le 1
\label{LGI1}
\end{eqnarray}
Using similar arguments one can derive an LGI for measurements at four different
times, $t_1$, $t_2$, $t_3$ and $t_4 = t_1 + 3\tau$ given by
\begin{eqnarray}
C_4 = C_{21} + C_{32} + C_{43} - C_{41} \le 2
\label{LGI2}
\end{eqnarray}
To avoid possible time-ordering ambiguities \cite{PRL2016}, we consider the symmetric
combination of the two-time correlation functions
\begin{eqnarray}
C_{ji} =  \langle \{ O(t_j), O(t_i) \} \rangle/2.
\label{Cji}
\end{eqnarray}
The anticommutator $\langle \{ O(t_j), O(t_i) \} \rangle/2 = \left( O(t_j) O(t_i) + O(t_i) O(t_j) \right)/2$ is
Hermitian \cite{NMAliChen,fritz}.

We investigate the dynamics of the Leggett-Garg inequality for a two-level system under spontaneous decay,
with the measurement operator $O=\sigma_x$ and the two-time correlators given by Eq.~(\ref{Cji}).
Consequently, the two-time correlation function $C_{21}$ is given by
\begin{eqnarray}
\label{C21A}
C_{21} &=& \langle \sigma_x(t_2) \sigma_x(t_1) + \sigma_x(t_1) \sigma_x(t_2) \rangle /2 \\
\nonumber
&=& \frac{1}{2} \{ \langle \sigma_{+}(t_1) \sigma_{-}(t_2) \rangle + \langle \sigma_{-}(t_1) \sigma_{+}(t_2) \rangle \\
\nonumber
&&{} + \langle \sigma_{+}(t_2) \sigma_{-}(t_1) \rangle + \langle \sigma_{-}(t_2) \sigma_{+}(t_1) \rangle \}
\end{eqnarray}
as the two-time correlation functions $\langle \sigma_{-}(t_1) \sigma_{-}(t_2) \rangle$ and
$\langle \sigma_{+}(t_1) \sigma_{+}(t_2) \rangle$ vanish for any pair of time $t_1$ and $t_2$. Then
combining Eqs.~(\ref{ttc1}), (\ref{ttc2}) and (\ref{C21A}) we have
\begin{eqnarray}
\label{C21B}
C_{21} &=& {\textrm Re} \{ c_1(t_2) c_1^{\ast}(t_1) e^{-i\omega_0(t_2-t_1)} + \\
\nonumber
&& |c_0|^2 c_1(t_2-t_1) e^{-i\omega_0(t_2-t_1)} \},
\end{eqnarray}
since $\langle \sigma_{+}(t_2) \sigma_{-}(t_1) \rangle$ and $\langle \sigma_{-}(t_2) \sigma_{+}(t_1) \rangle$
are the complex conjugates of $\langle \sigma_{+}(t_1) \sigma_{-}(t_2) \rangle$ and
$\langle \sigma_{-}(t_1) \sigma_{+}(t_2) \rangle$ respectively.

\section{Physical realization, results and discussion}\label{sec:results}

\begin{figure}[h]
\centering{\rotatebox{0}{\resizebox{7.0cm}{5.5cm}{\includegraphics{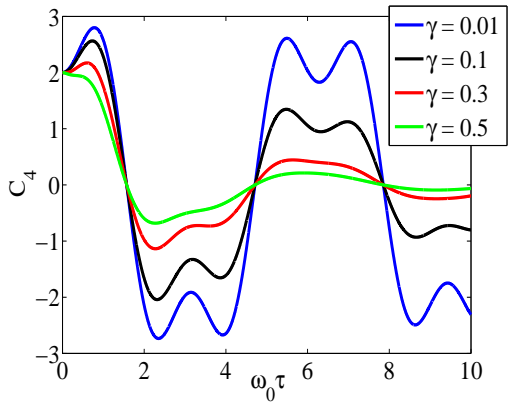}}}}
\caption{\label{fig1}  (Color online) We show the dynamics of Leggett-Garg inequality for
different system-environment coupling strengths. We plot $C_4$ as a function of $\tau$ for
four different values of $\gamma = 0.01 ~,~\gamma = 0.1 ~,~\gamma = 0.3 ~,~\gamma = 0.5$.
The other parameters are taken as $\omega_0t_1=0$, $\Lambda = 5 \omega_0$, and $\Delta=0$.}
\end{figure}

\begin{figure}[h]
\centering{\rotatebox{0}{\resizebox{7.0cm}{5.5cm}{\includegraphics{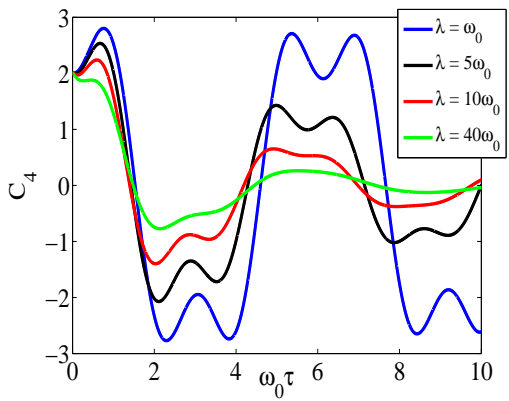}}}}
\caption{\label{fig2}  (Color online) We show the dynamics of Leggett-Garg inequality for
different cutoff frequency $\lambda$. We plot $C_4$ as a function of $\tau$ for
four different values of $\lambda = \omega_0~, ~\lambda = 5 \omega_0 ~,~\lambda
= 10\omega_0 ~,~\lambda = 40\omega_0$. The other parameters are taken as $\omega_0t_1=0$,
$\gamma=0.5$, and $\Delta=10\omega_0$.}
\end{figure}

\begin{figure}[h]
\centering{\rotatebox{0}{\resizebox{7.0cm}{5.5cm}{\includegraphics{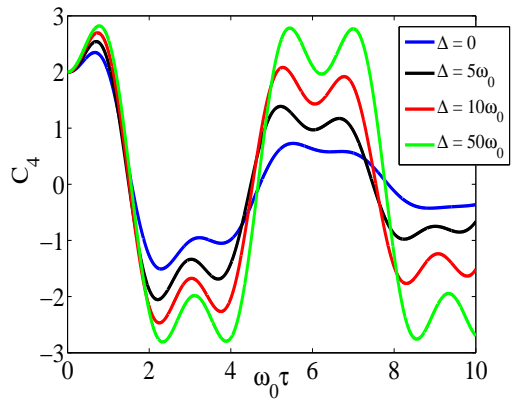}}}}
\caption{\label{fig3}  (Color online) We show the effect of varying detuning $\Delta$ on
the dynamics of Leggett-Garg inequality. We plot $C_4$ as a function of $\tau$ for
four different values of $\Delta = 0~, ~\Delta = 5 \omega_0 ~,~\Delta = 10\omega_0 ~,~\Delta = 50\omega_0$.
The other parameters are taken as $\omega_0t_1=0$, $\gamma = 0.2$, and $\lambda=5\omega_0$.}
\end{figure}

We consider Lorentzian spectral density of the environment which is widely used in the context of non-Markovian
open quantum systems recently \cite{Breuer02,Breuer09,Breuer16}
\begin{eqnarray}
J(\omega) = \frac{1}{2\pi} \frac{\gamma \lambda^2}{(\omega_0-\omega-\Delta)^2 + \lambda^2},
\label{spectra}
\end{eqnarray}
where $\gamma$ describes the coupling strength, $\lambda$ is the spectral width and $\Delta$ is the detuning.
For this $J(\omega)$, the exact probability amplitude $c_1(t)$ of Eq.~(\ref{amp1N}) can be solved analytically
\begin{eqnarray}
\!\!\!\!\!\!\!c_1(t) \!=\! c_1(0) e^{-\frac{1}{2} (\lambda-i\Delta)t}\!\!\left( \cosh \frac{d t}{2}
\!+\! \frac{\lambda - i \Delta}{d} \sinh \frac{d t}{2}  \right)
\end{eqnarray}
where $d=\sqrt{(\lambda-i\Delta)^2 - 2 \gamma \lambda}$. With the spectral density specified, the correlation
functions $C_{21}$, $C_{32}$, $C_{43}$, and  $C_{41}$ can then be calculated exactly from Eq.~(\ref{C21B}).
We show the exact dynamics of Leggett-Garg inequality, specifically we plot $C_{4}$ for a wide range of
system-environment parameters. The initial environment state is considered to be in the thermal equilibrium state and the system
is arbitrarily chosen to $\left\vert \Psi \right\rangle =\frac{1}{\sqrt{2}}\left(\left\vert + \right\rangle + \left\vert - \right\rangle \right)$,
hence $\rho_S (0) = | \Psi \rangle \langle \Psi |$. Here $|+\rangle$ and $|-\rangle$ are the eigenstates of $\sigma_x$.
In Fig.~\ref{fig1}, we show the dynamics of Leggett-Garg inequality as a function of time $\tau$ for different system-environment
coupling strengths $\gamma$ at a fixed cutoff frequency $\Lambda=5\omega_0$ and detuning $\Delta=0$. For simplicity,
we also set $\omega_0t_1=0$. Different curves represent different coupling strengths, namely $\gamma=0.01$ (blue), $\gamma=0.1$ (black),
$\gamma=0.3$ (red), and $\gamma=0.5$ (green). For weak system-reservoir coupling, the system shows nonclassical behavior
(violation of LGI) as a function of measurement intervals $\tau$. The violation of LGI is reduced and limited to a very short
measurement intervals $\tau$ as one increases the coupling strength, the nonclassicality of the open system eventually vanishes.
Next in figure \ref{fig2}, we show the dynamics of $C_{4}$ with different cutoff frequency $\lambda$. We plot $C_4$ as a
function of $\tau$ for four different values of $\lambda = \omega_0$ (blue), $\lambda = 5 \omega_0$ (black),
$\lambda = 10\omega_0$ (red), and $\lambda = 40\omega_0$ (green). The other parameters are taken as $\gamma=0.5$,
$\Delta=10\omega_0$, and $\omega_0t_1=0$. We observe (Fig.~\ref{fig2}) a reduced violation of Leggett-Garg inequality
as we increase the cutoff frequency of the spectral density. For higher values of $\lambda$, the system dynamics goes beyond
classical description (violation of LGI) for short measurement intervals $\tau$. In Figure \ref{fig3}, we examine the effect of
varying detuning $\Delta$ on the dynamics of Leggett-Garg inequality. The dynamics of $C_4$ is shown with four different
values of $\Delta = 0$ (blue), $\Delta = 5 \omega_0$ (black), $\Delta = 10\omega_0$ (red), and $\Delta = 50\omega_0$ (green).
The other parameters are taken as $\gamma = 0.2$, $\lambda=5\omega_0$, and $\omega_0t_1=0$. This indicate that an enhanced
nonclassicality or LGI violation when the reservoir spectral density is detuned ($\Delta \gg \omega_0$) from system frequency.
The LGI violation also depends on the initial time $t_1$ of the first measurement. We also have studied numerically the effect of
varying $t_1$ on the dynamics of Leggett-Garg inequality. It is observed that the nonclassicality of the open system will be wiped
out if we allow the system to evolve under the environment for a long time before performing the measurements.

For experimental investigation of the nonclassicality or quantumness through Leggett-Garg inequality, we propose
to consider a two level quantum emitter (a solid state qubit) positioned close to a two-dimensional metal-dielectric
interface \cite{Gonzalez10,Gonzalez14,GuangYin} keeping in mind the physical motivation to consider
Lorentzian spectral density. The quantum emitter coupled to the metal-surface electromagnetic
modes can be described by the Hamiltonian (\ref{TotH}), and the problem can be solved exactly using
Wigner-Weisskopf approach as discussed in Sec.~\ref{sec:model}. Dynamics of the excited-state population and
reversible coherent dynamics for this physical system was studied recently \cite{Gonzalez14} but our main focus
in this work is on two-time correlation functions and probing nonclassicality using Leggett-Garg inequality. The spectral
density of the metal-surface electromagnetic field is strongly modified in presence of the quantum emitter. A recent
research revealed \cite{Gonzalez14} that with small enough separation between the quantum emitter and the
metal-dielectric interface, the spectral density (which comprises information about the density of the surface
electromagnetic field, and also the coupling between quantum emitter and the metal surface) can take a form
of the Lorentzian distribution.

\section{conclusion}

In summary, we have used Leggett-Garg inequality as a nonclassicality witness for an open quantum system.
We investigate the dynamical loss of quantumness through Leggett-Garg inequality for a two level
system (qubit) spontaneously decaying under a general non-Markovian dissipative environment.
Our analysis is exact as we have calculated the two-time correlation functions exactly without using
Born-Markov approximations. We show the exact dynamics of Leggett-Garg inequality for a wide range of
system-environment parameters. Further experimental investigations are required to explore the nonclassicality
of open quantum systems using two-time correlation functions which are experimentally measurable.

\begin{acknowledgments}
M. M. Ali acknowledges the support from the Ministry of Science and Technology of Taiwan and
the Physics Division of National Center for Theoretical Sciences, Taiwan. P.-W. Chen would like to
acknowledge support from the Excellent Research Projects of Division of Physics, Institute of
Nuclear Energy Research, Taiwan.
\end{acknowledgments}

\end{document}